\documentclass[reprint,aip,cha]{revtex4-1}
\usepackage{graphicx}

\begin{document}

\title[]%A {THz}-Vibration to {THz}-Radiation Converter Based on Gold Nanoobjects]
{A {THz}-Vibration to {THz}-Radiation Converter Based on Gold Nanoobjects:\\
 a Feasibility Study\bigskip}
\author{Kamil Moldosanov}
\address{Kyrgyz-Russian Slavic University,
44 Kiyevskaya St., Bishkek 720000, Kyrgyzstan}
\email{altair1964@yandex.ru}
\author{Andrei Postnikov}
\address{Universit\'e de Lorraine, 
LCP-A2MC, 1 Bd Arago, F-57078 Metz, France}
\email{andrei.postnikov@univ-lorraine.fr}

\begin{abstract}
The estimations done confirm a feasibility of the idea that gold nanobars (GNBs)
and nanorings (GNRs) irradiated by microwaves could become the terahertz (THz)
emitters with photon energies within the full width at half maximum
of longitudinal acoustic phononic DOS of gold (${\approx}\,13.7-17.5$~meV,
i.e., $3.3-4.2$~THz). A scheme of the THz radiation source is suggested
based on the domestic microwave oven irradiating a substrate
with multiple deposited GNBs or GNRs.
\end{abstract}

%\begin{keyword}THz emitter, nanobar, nanoring,
%longitudinal acoustic phonon, microwave photon
%\end{keyword}
\maketitle

%\section{} %{Introduction}

\noindent
The terahertz (THz) range of the electromagnetic waves, the intermediate one
between microwaves and infrared (IR), is often discussed in reference to
the ``terahertz gap'', where ``electronics meets optics''. The corresponding radiation,
exhibiting properties common to one or the other of its neighbouring ranges, can be
refracted and focused by lenses, like IR rays, and penetrate many optically opaque 
barriers, like microwaves. However, the methods of radiation generation and detection,
elaborated for these adjacent ranges, are not quite efficient in THz and do face
grave challenges. This hinders the creation of devices which can be a priori expected
to open new fields of application, notably in medicine (non-invasive early diagnosis
of cancer) and security (detection of concealed goods). The development of novel
THz generation and detection methods, not borrowed from bordering technologies,
invariably attracts the attention of the THz community.

\begin{figure}[b] % --------------------------------------------
\includegraphics[width=0.47\textwidth]{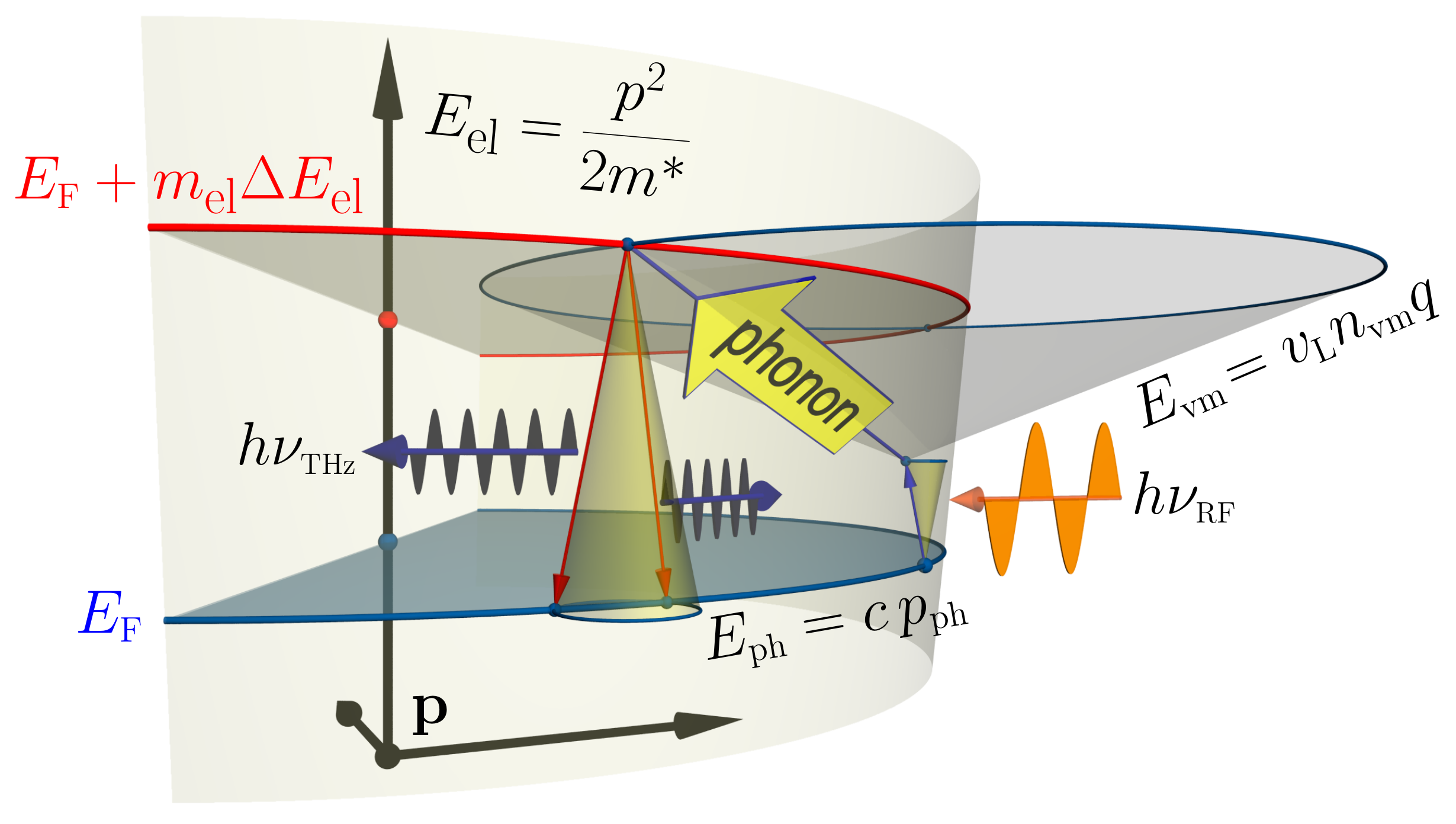}
\caption{\label{fig:3d-schema}
Scheme of absorption of microwave radiation $h\nu_{\rm RF}$ by an electron
at $E_{\rm F}$, assisted by an absorption of a longitudinal phonon,
with subsequent emission of a terahertz quantum $h\nu_{\rm THz}$. 
The corresponding dispersion laws are indicated.
See text for details.}
\end{figure}

Alternating electric fields of the THz frequency range are well known to exist
in solids: such are the fields created by longitudinal acoustic vibrational modes
(LAVMs). How can one ``extract'' them from a solid and transform into the radiative 
energy of the THz range? It turns out that in metal nanoparticles, the LAVMs can be
converted into electromagnetic waves with the help from the Fermi electrons. Our idea
consists in simultaneous excitation of these latter by exposing them to a THz field
of longitudinal phonon and the microwave radiation. This should occur in
a nanoparticle, where the conditions exist for an excited Fermi electron
to relax fast enough via emitting a THz quantum. Specifically, gold nanobars (GNBs)
and gold nanorings (GNRs) seem to be good candidates for implementing this idea.
In the following, we discuss the theoretical prerequisites for the effect
within the quasiparticle approach. The microwave photons are considered to be
quasiparticles with the dispersion law $E_{\rm ph}=c{\cdot}p_{\rm ph}$
($c$ is the speed of light); the longitudinal phonons would be quasiparticles
with dispersion law $E_{\rm vm}=v_{\rm L}{\cdot}n_{\rm vm}{\cdot}q$, where
$v_{\rm L}$ is the (longitudinal) sound velocity in gold; finally, the electrons
would be quasiparticles with the dispersion law $E=p^2/(2m^{\ast})$, 
$m^{\ast}$ being the electron effective mass. 
These three quasiparticle species may interact, subject to their specific quantisation
conditions and the energy / momentum conservation laws. The latter are grasped
by a schema in Fig.~\ref{fig:3d-schema} where an excitation of
a Fermi electron ``upwards'' through $m_{\rm el}$ quantization steps
${\Delta}E_{\rm el}$ is brought about by a joint absorption of a RF quantum
and an appropriate longitudinal phonon. Subsequently the excited electron
relaxes to the Fermi energy, releasing a quantum at THz.
In the following, we argue about conditions under which the THz-emitting
channel of relaxation may precede other imaginable mechanisms, e.g.,
that releasing the extra energy into the phonon bath again, the case we
considered in a related but different context \cite{FANEM2015}. It is essential
for our analysis that the nanoobjects are much larger in one dimension
than in other two, hence having the shape of nanobars or closed thin nanorings.

 % -------------------------------------------------

\begin{figure*}[t] % --------------------------------------------
\centerline{\includegraphics[width=0.80\textwidth]{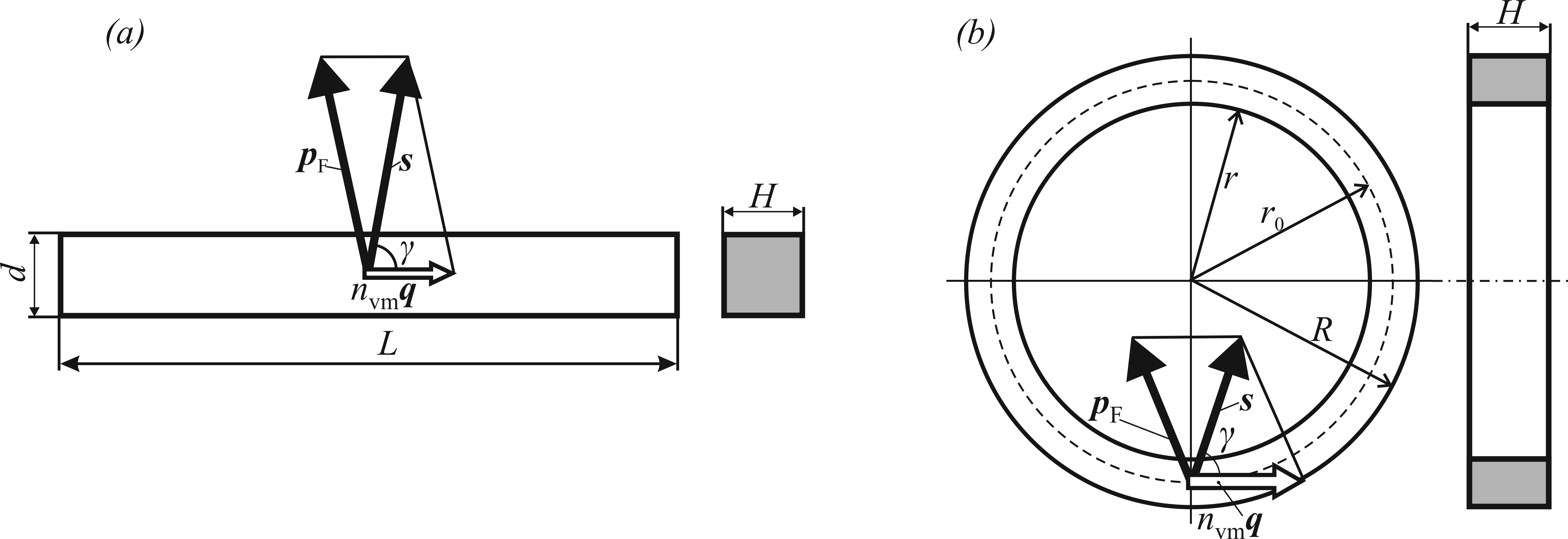}}
\caption{\label{fig:objects}
Geometry of nanoobjects and phonon / electron momenta in GNB (a) and in GNR (b).
For simplicity, the momentum $\mathbf{p}_{\rm ph}$ of the microwave photon is omitted,
by force of relations $p_{\rm ph}{\ll}n_{\rm vm}q$, $p_{\rm ph}{\ll}p_{\rm F}$.
Note that the compression wave may travel in both directions along the bar or the ring,
whereas the propagation of excited electrons is confined along the conical surfaces
with large opening angle, $2{\gamma}\,\approx\,144.6^{\circ}$.}
\end{figure*} % -----------------------------

We start with some numerical estimates.
The strongest ``assistance'' of longitudinal phonons to electron excitations
is expected within the full width at half maximum of a pronounced peak
in the density of longitudinal acoustic modes of gold 
(energies $\sim\,13.7-17.5$~meV, i.e. frequencies $\sim\,3.3-4.2$~THz) --
see, e.g., Mu\~noz \emph{et al.}\cite{PRB87-014301}.
For the microwave irradiation, a standard domestic microwave oven would offer
a simple practical source at $\nu_{\rm RF}=2.45$~GHz, i.e., 
$h\nu_{\rm RF}\,\approx\,1.01{\cdot}10^{-2}$~meV. As this is much smaller
than the above phonon-related values, 
the peak outcome of the THz radiation is not expected to be shifted 
from the peak of the phonon density of modes, i.e., around 3.77~THz (see below).

The sizes of GNBs and GNRs have to be much smaller than the skin depth in gold
at 2.45~GHz, that makes ${\approx}1.5$~$\mu$m. This justifies
considering the Fermi electrons in the following as free ones.

Fig.~\ref{fig:objects} sets the geometry of GNBs (left panel)
and GNRs (right panel), for the discussion that follows.
In GNBs, the compression waves (longitudinal phonons with the momentum
$n_{\rm vm}\mathbf{q}$ and the energy 
$E_{\rm vm}=v_{\rm L}{\cdot}n_{\rm vm}{\cdot}q$) propagate along its length.
It will be argued below that
the momentum of an excited electron stands at a large angle $\gamma$ to that
of the longitudinal phonon; however, the electron cannot leave the GNB
because of substantial work function of gold 
($\approx\,4.3$~eV)\cite{AshMerm_book}.
On creating conditions to prevent the scattering of the excited electron at the GNB
boundaries and its relaxation via emission of a longitudinal phonon, the setup
will be given for channeling the relaxation process into an emission of a THz photon.

In GNR (Fig.~\ref{fig:objects}b), the compression waves run along the cyclic contour, 
to which the phonon momentum $n_{\rm vm}\mathbf{q}$ is tangential.
Similarly to the case of GNB, the momentum of an excited electron stands
at large angle $\gamma$ to the momentum of the phonon.

The absorption condition for the microwave-range photon is the following:
\begin{equation}
m_{\rm el}\,{\Delta}E_{\rm el} = n_{\rm vm}\,{\Delta}E_{\rm vm}
+h{\nu}\,\approx\,n_{\rm vm}\,{\Delta}E_{\rm vm}\,,
\label{eq:01}
\end{equation}
where $m_{\rm el}$ is the number of electron energy ``gaps''
(quantization steps between confinement-induced discrete energy levels in nanoobject),
and $n_{\rm vm}$ is a number of vibrational quanta (for LAVMs). According to
the Kubo formula\cite{JPSJ17-975,JPhysColloq38-C2-69}
(see also the Appendix in Ref.~\onlinecite{FANEM2015}),
the step in the electron energy levels ${\Delta}E_{\rm el}$ relates
to the number of gold atoms $N_{\rm a}$ as follows:
${\Delta}E_{\rm el}\,\approx\,(4/3){\cdot}(E_{\rm F}/N_{\rm a})$. 
Then the condition (Eq.~\ref{eq:01}) that $m_{\rm el}$ energy steps must embrace
$n_{\rm vm}$ vibrational quanta takes the form:
\begin{equation}
m_{\rm el}{\cdot}(4/3){\cdot}(E_{\rm F}/N_{\rm a})\;\approx\;
n_{\rm vm}\,v_{\rm L}\,h/L\,,
\nonumber
\end{equation} 
where $L$ is the confinement length delimiting the propagation of compression waves
(i.e., the maximal dimension of a GNB or the median circumference of a GNR,
correspondingly), and the sound velocity $v_{\rm L}$ relates frequency to wave vector.
Further on, assuming that the density in a GNB equals that of bulk gold,
the number of atoms in GNB can be expressed via density of gold $\rho$,
atomic mass $m_{\rm a}$ and the nanoobjects's volume $V$. The resulting
$m_{\rm el}/n_{\rm vm}$ ratio for the two cases is as follows: 
\begin{equation}
\frac{m_{\rm el}}{n_{\rm vm}}\,\approx\,\frac{3}{4}\,
\frac{\rho\,v_{\rm L}\,h}{E_{\rm F}\,m_{\rm a}}{\cdot}\!
\left\{\!\!\begin{array}{l@{\mbox{\hspace*{0.4cm}}}l}
dH & \mbox{(case of GNB)\,;} \\
(R-r)H & \mbox{(case of GNR)\,.}
\end{array}\right.
\nonumber
\end{equation}
The numerical values for gold $E_{\rm F}=5.53$~eV\cite{AshMerm_book},
$m_{\rm a}=197$ in atomic units, or $\approx\;3.27{\cdot}10^{-22}$~g,
$\rho=19.3$~g/cm$^3$, $v_{\rm L}=3.23{\cdot}10^5$~cm/s,
(see, e.g., Ref.~\onlinecite{IJAPM3-275})
yield, for both $dH$ (case GNB) and $(R-r)H$ (case GNR)
the value of $\approx\;9.35{\cdot}(m_{\rm el}/n_{\rm vm})$.

To provide high intensity of THz emission, the number of quantizied energies
within the FWHM of the longitudinal phononic density of modes ought to be large enough.
In order to increase the efficiency of conversion, each ``allowed'' mode
within the FWHM has to be used, that imposes $(m_{\rm el}/n_{\rm vm})$ to be
an integer. On the other hand, an average excited electron should not be scattered
on its way towards the GNB boundary. Hence, $d/2$ and $H/2$ should reasonably
not exceed the electron mean free path $l_0$ in gold nanoparticles. Rough estimates
in our earlier work yield $l_0\,\approx\,1.7$~nm,\cite{FANEM2015}
that reveals the ``working'' matching relation $(m_{\rm el}/n_{\rm vm})=1$
and not larger. Specifically, accepting this relation and
$H\,\sim\,3.1$~nm, we get 
\begin{eqnarray}
d &{\sim}& 3~\mbox{nm (for GNB), or} \nonumber \\
(R-r) &{\sim}& 3~\mbox{nm (for GNR).}
\label{eq:02}
\end{eqnarray}

Each ``allowed/working'' phonon can promote the Fermi electron onto the vacant level
corresponding to this phonon's energy, measured from $E_{\rm F}$. On relaxation,
the energy of the emitted THz photon would roughly match (neglecting a much smaller
energy of the microwave photon) that of LAVM. Consequently, the ``primary'' emitted
THz spectrum (intensity distribution over energies) is expected to roughly follow
the distribution of the LAVM density in gold. However, the resulting spectrum
issued by the device would be limited by the transmission width of the band
resonance filter (see below in Fig.~\ref{fig:oven}, pos.~4). 

In addition to the energy conservation law for GNBs / GNRs expressed
by (Eq.~\ref{eq:01}), we take now into account the momentum conservation,
making reference to the schemes depicted in Fig.~\ref{fig:objects}.
The LAVM with momentum $n_{\rm vm}\mathbf{q}$ propagates along the GNB's length
(or, in GNR, over the median cycle contour along the ring). Everywhere within
the GNB / GNR, a Fermi electron with momentum $\mathbf{p}_{\rm F}$ can be encountered,
prone to absorb a microwave photon. The latter's contribution $p_{\rm ph}$
to the electron momentum will be neglected here due to its relative smallness
(as was done above for the energy $h{\nu}$). Assuming the vector $\mathbf{s}$
to be the sum of $\mathbf{p}_{\rm F}$ and $n_{\rm vm}\mathbf{q}$,
the following relation holds:
\begin{equation}
p_{\rm F}^2 = s^2 + (n_{\rm vm}q)^2-2s(n_{\rm vm}q)\cos\gamma\,,
\nonumber
\end{equation}
whence the angle that defines the propagation direction of the excited electron
reads:
\begin{equation}
\gamma=\arccos\frac{s^2 + (n_{\rm vm}q)^2-p_{\rm F}^2}
{2s(n_{\rm vm}q)}\,,
\nonumber
\end{equation}
where $s=\sqrt{2m(E_{\rm F}+n_{\rm vm}{\Delta}E_{\rm vm})}$. On expressing
the vibration mode's momentum and energy quanta, $q$ and ${\Delta}E_{\rm vm}$
respectively, in terms of the length of path along which the compression waves
propagate, $L$, and the longitudinal sound velocity $v_{\rm L}$, 
\begin{equation}
q = h/L \quad\mbox{and}\quad {\Delta}E_{\rm vm}=v_{\rm L}\,h/L\,,
\nonumber
\end{equation}
one arrives at the estimate for $\gamma$ in terms of $n_{\rm vm}$ and $L$:
\begin{equation}
\gamma = \arccos\frac{m\,v_{\rm L}+n_{\rm vm}h/(2L)}
{(p_{\rm F}^2+2m\,n_{\rm vm}h\,v_{\rm L}/L)^{1/2}}\,.
\nonumber
\end{equation}
Note that in GNB, $L$ is the nanobar's length, whereas in GNR,
the median contour length $\pi(R+r)$ should be taken instead
(see Fig.~\ref{fig:objects}b), hence 
\begin{equation}
\gamma = \arccos\frac{m\,v_{\rm L}+n_{\rm vm}h/[2\pi(R+r)]}
{\{p_{\rm F}^2+2m\,n_{\rm vm}h\,v_{\rm L}/[\pi(R+r)]\}^{1/2}}\,.
\nonumber
\end{equation}
This yields, taking into account the condition 
$n_{\rm vm}\,{\Delta}E_{\rm vm}+h{\nu}\,{\approx}\,15.6$~meV along with
numerical values of gold parameters and constants: for both GNB and GNR,
$\gamma\,{\approx}\,\arccos(0.3038)\,{\approx}\,72.3^{\circ}$. 

The excited electron transverses the GNB / GNR at this angle to the line
of phonon propagation and reaches the boundary. The ensuing imaginable scenarios are:
$(i)$ elastic scattering back into the GNB / GNR volume, that would simply ``postpone''
the electron's relaxation; 
$(ii)$ releasing the ``originally borrowed'' longitudinal phonon and
the microwave photon, restoring the initial situation; 
$(iii)$ inducing just a longitudinal phonon of slightly higher energy,
ultimately translating the initially absorbed radiation into heat; 
$(iv)$ releasing the surplus energy into a transversal surface phonon; 
$(v)$ emitting the surplus energy in form of a THz photon. We emphasize that
the electron with the given excitation energy won't exit the sample because
the work function of gold is too high to overcome. Let us briefly discuss these
possibilities. The cases $(ii)$ and $(iii)$ will be unfavoured (and hence rare)
by force of geometry considerations, in view of nearly normal orientation of
the electron's momentum to the surface. 
Even as the case $(iii)$ would mean an ``useless'' (in the sense of our objective)
waste of the microwave energy, it will ``populate'' the system with longitudinal
phonons, thus enhancing the possibility of scenario $(v)$ and compensating
for the ``cooling'' implied by the latter. Note that this very process came about
as the useful one in a different context, in our work on hyperthermia
with the help of gold nanoparticles, Ref.~\onlinecite{FANEM2015}.
The case $(iv)$ might seem plausible 
(albeit similarly ``useless''); the consideration against it is that a transversal
phonon will likely be ``out of resonance'' with the energy delivered by a longitudinal 
phonon, since their dispersion relations are different, and the confinement-imposed
energy quantization would likely prevent the necessary energy match of their respective
$n_{\rm vm}{\Delta}E_{\rm vm}$ values. Putting it differently, the major peaks
in the densities of modes of transversal and longitudinal phonons in gold
(and particularly in gold nanoparticles) are well separated, as can be seen,
e.g., in Figures 7 and 8 of Ref.~\onlinecite{FANEM2015}. This leaves us
with the scenario
$(v)$, the only ``useful'' one, at which the electron relaxes back to the Fermi level
via a radiative transition emitting photon at $\approx\,3.77$~THz, i.e. with
the energy $\approx\,15.6$~meV (or close to it, according to that of
the ``primary'' photon).

In order to reduce the probability of the electron's elastic scattering at the GNB
boundary, and hence to enforce the decay probability into THz photon emission, one can
try to enhance the electron density of states at the Fermi level. This can be
achieved, e.g., by ion implantation of Ta or Fe impurities into the subsurface layer
of nanoobjects facing the lens (Fig.~\ref{fig:oven}). The idea is, the doping of
bulk gold with Ta or Fe atoms create impurity $d$-levels (in one of spin channels,
since the impurities are magnetic) at the Fermi energy \cite{JNP6-061709}.
Should the scheme work, the GNB / GNR array would turn into a steady source
of THz-range quasicontinuous spectrum centered at 3.77~THz, or 15.6~meV
(the major peak in the LAVM of gold) and possessing the FWHM of the latter
($\approx\,0.9$~THz, or 3.74~meV). ``Quasicontinuous'' (in the frequency domain)
implies here that the emitted THz spectrum is expected to roughly follow the phonon
density of modes, however, in consistency with the confinement conditions imposed on
the nanoobject, i.e., the discreteness of the frequency of the order of
$v_{\rm L}/L\,\sim\,10^{10}$~Hz (see the discussion about the practical values of $L$
below). ``Steady'' means that the radiation is emitted continuously in the time domain,
as long as the microwave oven operates, and not in the pulsed regime. 

The above estimates of the nanoobjects' size (Eq.~\ref{eq:02}) concern the
``thickness'' \mbox{$(d)\,/\,(R-r)\,/\,(H)$} but neither the GNB's length $L$
nor the GNR's mean radius $r_0=(R+r)/2$. However, these dimensions come into
the foreground as
we think over how to reduce the heating of nanoparicles by 2.45~GHz radiation, that
would occur due to uncertainty in the momentum value of the Fermi electrons. For sizes
as given in (Eq.~\ref{eq:02}), the uncertainty in the Fermi electron momentum
guarantees
that the momentum conservation will hold. In order to prevent the ``direct''
absorption of the microwave radiation by the nanoobjects, one has to keep the relevant
energy intervals incompatible -- e.g., by choosing the nanoparticles' sizes such that
the energy step ${\Delta}E_{\rm el}$ between the corresponding energy levels in 
the electron system would be by far larger than $h{\nu}=1.01{\cdot}10^{-5}$~eV.
A reference to the abovementioned Kubo formula\cite{JPSJ17-975,JPhysColloq38-C2-69}
yields the ``critical''  (in this sense) length dimensions (in nm) of GNBs:
$L<1.24{\cdot}10^4{\cdot}d^{-1}{\cdot}H^{-1}$,
or of GNRs: \mbox{$r_0<1.97{\cdot}10^3{\cdot}(R-r)^{-1}{\cdot}H^{-1}$}.
For example, for $d{\sim}3$~nm, $H{\sim}3.1$~nm from (Eq.~\ref{eq:02}), 
the maximal linear size $L_{\rm max}{\sim}1.33{\cdot}10^3$~nm. By choosing
$L\,{\approx}\,100$~nm, we can safely avoid a ``direct'' heating of GNBs
by microwave radiation. In case of GNRs, the choice $R-r{\sim}3$~nm,
$H{\sim}3.1$~nm yields the maximum mean radius 
$(r_0)_{\rm max}{\sim}2.11{\cdot}10^2$~nm.
A choice of radius of say $r_0=20$~nm, would securely help from a ``direct'' heating.

\begin{figure}[t] % ---------------------------------------------
\centerline{\includegraphics[width=0.38\textwidth]{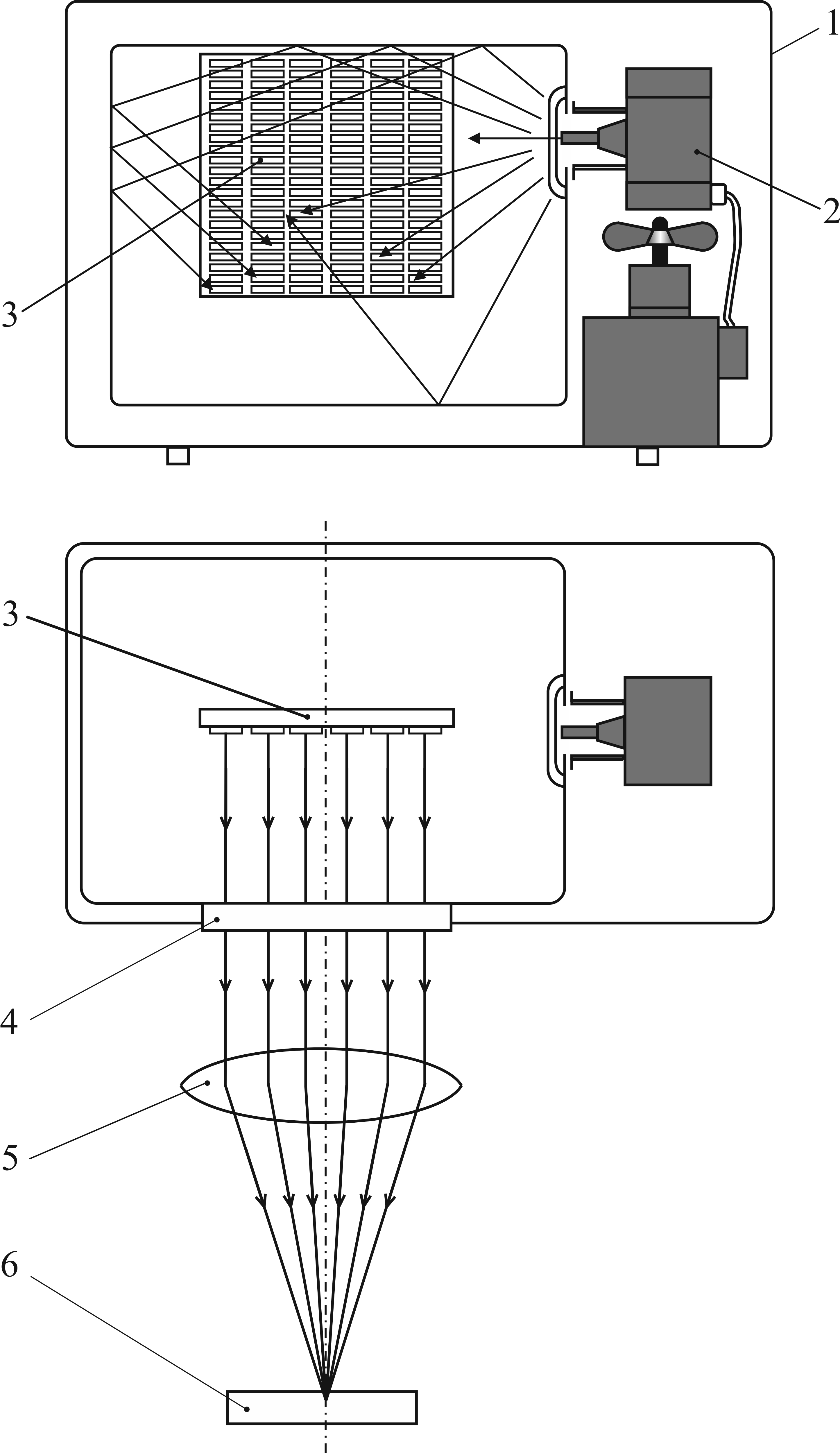}}
\caption{\label{fig:oven}
An outline of the THz radiation source based on the domestic microwave oven.
1:~microwave oven, 2:~magnetron ($\nu\,=\,2.45$~GHz), 
3:~substrate with densely deposited GNBs or GNRs, 
4:~THz band resonance filter ($\nu_0\,=\,3.77$~THz),
5:~lens, 6:~object to be studied.} 
\end{figure} % ----------------------------------------------------

The suggested idea of transforming THz ``sound'' into THz electromagnetic waves
can hopefully be realized as a source of continuous THz radiation; a possible scheme
is depicted in Fig.~\ref{fig:oven}. As a practical source of microwave radiation,
a magnetron of a domestic microwave oven is suggested, that operates at
$\nu\,=\,2.45$~GHz. In this case,
$h{\nu}\,{\approx}\,1.01{\cdot}10^{-2}$~meV$\,{\ll}\,15.6$~meV,
so that the approximation $h{\nu}\,{\ll}\,n_{\rm vm}{\Delta}E_{\rm vm}$ holds
and the approximation (\ref{eq:01}) is valid. A substrate with multiple GNBs
is placed into the oven chamber, where the GNBs are exposed to both direct irradiation
from the magnetron and that reflected from the chamber walls. Effectively,
the microwave photons could be absorbed anyway inside the GNBs, where consequently
the THz photons will be emitted. The resulting THz radiation is channeled out of
the chamber via a THz band resonance filter with resonance frequency 
$\nu_0\,=\,3.77$~THz ($h{\nu}_0\,=\,15.6$~meV), whereby the scattered microwave 
radiation is retained. The outcoming THz waves will be focused by a lens
(fabricated, e.g., from high-density polyethylene).

Finally, two technical remarks could be in place. The first one concerns
the validity of a quasiparticle picture implied above for the description of electrons
in nanoobjects: are the latter ``large enough'' for the discussion in terms of
the Fig.~\ref{fig:3d-schema} to make sense? The Fermi radius in (bulk) gold
being $k_{\rm F}\,\approx\,1.20{\cdot}10^{10}$~m$^{-1}$, the de Broglie wavelength
of an excited electron measures $\sim\,0.52$~nm, that fits, on the average,
$\sim\,3$~times along the free path to the nanoobject's boundary. Hence
the situation is just at the border of applicability region for the quasiparticle
approximation; we use the latter out of practical convenience, keeping in mind
the care to be taken.

The second remark is that the above estimated dimensions of GNBs and GNRs are
on the borderline in yet another way, namely, their size is at the resolution limit
for available lithographic techniques (intrinsic resolution of the electron beam
lithography \cite{ASS164-111} is $\sim\,3-5$~nm). Presumably, it is 
technologically simpler to fabricate the GNBs than the GNRs. A number of strategies
exists to produce gold nanorods and nanowires of $\sim\,3$~nm diameter
(see, e.g., Refs.~\onlinecite{NanoLett2-793,Langmuir24-9855,ACSNano2-1939})
which could be used as the GNBs. These strategies, possibly, could be tailored
for making the GNBs on the substrate or embedded into matrix transparent
in THz wavelength range. As for GNRs, 
Tseng \emph{et al.} \cite{Nanotechnology21-295102}
fabricated nanorings using the colloidal lithography process; hopefully, it could be
tailored to provide the GNR sizes estimated here.

Summarizing, fully aware that extremely small size nanoobjects are difficult
to fabricate, we offer brief assessments hinting that the conversion of the THz
lattice vibrations to THz electromagnetic radiation could be possible via 
electron excitation / relaxation processes within the GNBs or GNRs. On the basis
of the principles outlined, a practical source of THz radiation might be hopefully
created, which would find applications, e.g., for the study of biological objects
(cancer diagnostics) and / or for detection of concealed objects.

\section*{Acknowledgements}
\noindent
We thank Dr. Mu{\~n}oz and Dr.~Kresch for useful discussions
of their measurements of the phonon spectrum in gold.

%\bibliography{GNP}
%\end{document}

%\section*{}%Bibliography}

\providecommand{\latin}[1]{#1}
\providecommand*\mcitethebibliography{\thebibliography}
\csname @ifundefined\endcsname{endmcitethebibliography}
  {\let\endmcitethebibliography\endthebibliography}{}

\end{document}